# Microscope calibration using laser written fluorescence

Alexander D. Corbett,[1,*] Michael Shaw,[2,5] Andrew Yacoot,[2] Andrew Jefferson,[3] Lothar Schermelleh,[3] Tony Wilson,[4] Martin Booth,[4] and Patrick S. Salter[4]

[1]*Department of Physics and Astronomy, University of Exeter, EX4 4QL, UK*
[2]*National Physical Laboratory, Hampton Rd, Teddington, TW11 0LW, UK*
[3]*Micron Oxford Advanced Bioimaging Unit, Department of Biochemistry, University of Oxford, South Parks Road, Oxford OX1 3QU, UK*
[4]*Department of Engineering Science, University of Oxford, Parks Road, Oxford, OX1 3PJ, UK*
[5]*Department of Computer Science, University College London, London, WC1 6BT, UK*
[*]*a.corbett@exeter.ac.uk*

**Abstract:** There is currently no widely adopted standard for the optical characterisation of fluorescence microscopes. We used laser written fluorescence to generate two- and three-dimensional patterns to deliver a quick and quantitative measure of imaging performance. We report on the use of two laser written patterns to measure the lateral resolution, illumination uniformity, lens distortion and colour plane alignment using confocal and structured illumination fluorescence microscopes.

## 1. Introduction

Fluorescence microscopes are essential tools across a wide range of scientific research. The greater repeatability of sample preparation and specificity of fluorescent labels have led to an increased emphasis on extracting quantitative data from images rather than making qualitative observations [1]. A further drive towards standardisation is the need to be able to make direct comparisons between data sets captured from a wide variety of microscope configurations. Whilst standards have been proposed in coherent microscopy [2], there is no single calibration standard that has been widely adopted in fluorescence microscopy that is able to provide a measure of key imaging performance parameters.

Fluorescent beads with diameters smaller than the diffraction limit, provide a popular means of measuring the point spread function (PSF) across the field of view. However, without careful preparation there is the risk of bead clustering which can distort the measurement. Standard protocols for preparing bead samples have been suggested by others [3] but these are generally labour intensive to implement. A further limitation is that the sub-diffraction size of the bead often results in (i) a weak fluorescence signal with low signal-to-background ratio and (ii) only a few sampling points across the PSF. Both of these factors contribute to a high degree of error in the PSF measurement. Finally, as the beads themselves are distributed randomly across the field of view, it is not possible to extract any measures of image distortion across the field of view.

There is now a pressing need for a tool which is able to provide accurate calibration information, allowing end users to have confidence in the integrity of their image data and the results derived from them. In recent years several commercial calibration standards have become available. Argolight use a laser to write features by the coalescing of metallic nanoparticles distributed throughout a glass substrate [4,5]. The features are stable with a broadband emission spectrum. However, the cost of the raw materials is significant and this has limited uptake by a research community that anticipates an inexpensive solution. GATTAquant have provided a customisable DNA-origami ruler to allow the adhesion of user-defined fluorophores at well-defined separations along a rigid strand of DNA [6]. Whilst these are provided at a fraction of the cost of an Argolight standard, the DNA strands themselves are distributed randomly across the field of view, limiting the number of imaging performance parameters that can be extracted from the images.

In this paper we demonstrate the utility of an inexpensive calibration standard that provides quantitative measures of the imaging performance in fluorescence microscopes. Direct laser writing with ultrashort pulses [7] is used to fabricate structures with three-dimensional resolution inside a polymer substrate, creating bright fluorescent features against a dark background. These features are used to construct a variety of fluorescent patterns embedded within the plastic to quantify different metrics of imaging performance. With a microscope slide form factor, the calibration patterns can be used quickly and easily, thereby reducing the barriers to its adoption by the research community.

## 2. Methods

### 2.1 Laser written fluorescence

It has been previously shown that when infrared pulses are tightly focused into glass or polymer substrates, microscopic voids are created [8–10]. Material within a thin region of the void interface is densified and shows increased autofluorescence [9,11]. In contrast to previous approaches which used laser fabrication to selectively bleach a dye bound within a plastic substrate [12], here we employ this enhanced autofluorescence to laser write patterns of fluorescent features for microscope calibration.

Fluorescent patterns were fabricated using direct laser writing with adaptive optics to compensate for aberrations generated at the polymer-glass interface. The details of the adaptive optics enabled laser writing set up can be found in [13]. In brief, the output beam from an amplified Ti:Sapph laser (790 nm, 250 fs, 1 kHz rep rate, Solstice, Spectra-Physics) was intensity modulated by a half-wave plate in a motorised mount. The intensity modulated beam was then expanded and relayed onto a spatial light modulator (SLM, Hamamatsu X10468-02) which was in turn imaged in a 4f configuration onto the back focal plane of the fabrication objective lens (Olympus 60X 1.4 NA). The SLM provided phase compensation for any residual aberrations (predominantly spherical) arising from the interface between the glass coverslip and polymer substrate. The corrections were made using a sensorless adaptive optics approach in a Zernike basis, with a metric related to the threshold for laser fabrication [14]. The necessary aberration correction closely matched that predicted by theory for spherical aberrations arising from a planar interface between media of different refractive index [15].

The calibration samples were constructed as shown in Fig. 1(a) with a polymer substrate sandwiched between a microscope slide and a coverslip. The polymer comprised a two-part epoxy mixture which was experimentally found to give optimum fluorescence signal-to-background ratio for the laser written features. The resin is hardened *in situ*, such that no mounting medium is required. Plastic adhesive spacer rings provided a repeatable substrate thickness of ~200 µm. During fabrication, the lateral and axial position of the sample is controlled by a high precision 3D translation stage (Aerotech ABL10100) relative to the fixed fabrication objective.

When writing lines of fluorescent features inside the polymer, the pulse spacing was determined by the lateral speed of the stage and the repetition rate of the laser. Typically, a lateral scan rate of 0.5 mm s$^{-1}$ and 1 kHz repetition rate is used to obtain a pulse spacing of 0.5 µm, such that the fluorescent features created by each pulse do not significantly overlap. The exact size of the fluorescent feature around the focal point of the laser is determined by the laser pulse energy, typically set in the range 3-8 nJ per pulse (i.e. 3-8 µW average power). In this way a square grid on a 10 µm pitch covering a region of 0.5 mm x 0.5 mm can be written (Fig. 1(b)). Each individual grid line is composed of four rows of fabricated features, giving the line a total width of 2 µm. An image of the grid pattern was processed to determine lens distortion and illumination uniformity.

As well as continuous fabrication at the repetition rate of the laser, single pulses are used to fabricate isolated features within the substrate. To create a three-dimensional array of 8 × 8 × 3 (X Y Z) isolated features on a 10 µm pitch, each feature was fabricated using five pulses of 7.4 nJ (Fig. 1(c)). This 3D array of points was used to determine the accuracy and precision with which it was possible to place individual features within the substrate. Additionally, single features were fabricated using multiple pulses and a range of pulse energies to validate the lateral resolution measurements made from the fluorescence images.

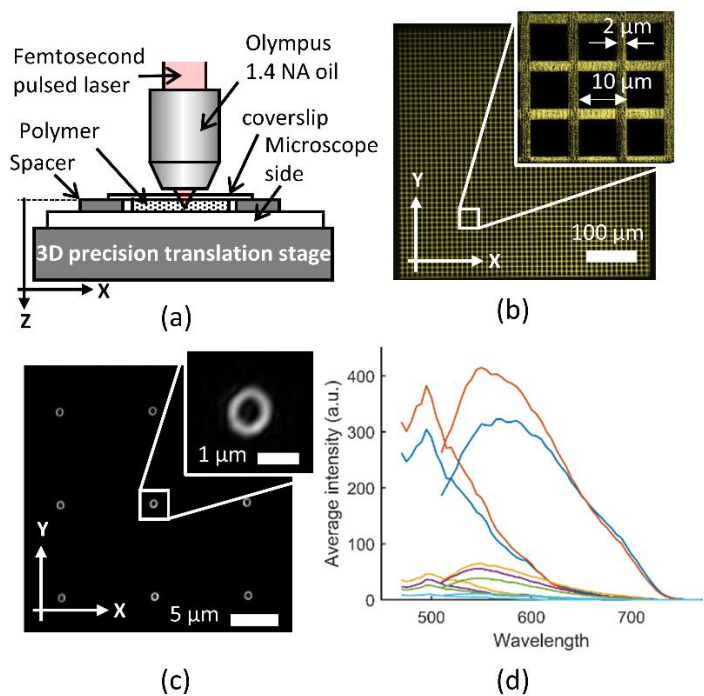

Fig. 1. (a) Fabrication of the features within the polymer substrate using a pulsed IR laser. (b) The fluorescent grid target shownning with the microstructure from the laser writing process, and magnified in the inset. (c) SIM images of individual fabrication features on a 10 µm pitch. (d) Excitation and emission fluorescence spectra for the laser fabricated regions. Wavelength is in nm. The fabrication powers used to produce the spectra are (according to line colour) orange = 8.0 nJ, dark blue = 6.8 nJ, yellow = 5.7 nJ, purple = 4.7 nJ, green = 3.8 nJ, light blue = background fluorescence.

*2.2 Excitation and emission spectra*

Key to the broad applicability of the fluorescent target is the relatively broadband emission profile of the fluorescence (Fig. 1(d)). This ensures compatibility with the majority of fluorescence filter sets and allows the target to be used to characterise scaling, chromatic aberration and other non-common path errors between detection channels. The excitation and emission spectra were measured using a Leica TCS SP5 X confocal laser scanning microscope. The supercontinuum laser on the setup provided an excitation range from 470-670 nm. When acquiring the excitation spectra, a single wavelength from the supercontinuum was used to excite the sample. The microscope was set to 'constant power' mode to compensate for the variable excitation brightness with wavelength in the supercontinuum source. The detection band started at a wavelength 30 nm longer than the excitation wavelength and was of fixed width (100 nm). The excitation wavelength was then stepped in 5 nm increments from 470 – 670 nm.

To acquire the emission spectrum, the sample is first excited at the peak excitation wavelength (495 nm). The fluorescent emission is then captured in a 5 nm band with the centre of the band covering a range from 10 nm longer than the excitation window (to avoid collecting excitation light) up to 780 nm. The excitation and emission spectra were collected for a number of different fabrication powers to indicate the scaling in fluorescence brightness with pulse energy (Fig. 1(d)).

*2.3 Spatial accuracy and precision*

To determine the spatial accuracy with which it is possible to fabrication a feature, a square array of 8 × 8 × 3 (X Y Z) features on a 10 µm pitch was created. Images of the array were recorded using a confocal microscope (FV1000, Olympus) whose transverse magnification and focusing stage had been calibrated against traceable dimensional standards using a grating and an optical interferometer respectively.

The displacement of the microscope's nosepiece focusing stage was calibrated using an NPL plane mirror differential optical interferometer [16]. Under ideal conditions, the uncertainty associated with the interferometer measurements is less than 1 nm; however in practice vibrations and thermal fluctuations within the laboratory significantly increased the achievable measurement uncertainty. The interferometer, along with an annular reference mirror, was mounted on the microscope's xy translation stage, which remained stationary throughout the calibration. A second plane (measurement) mirror was mounted in the microscope objective turret on the focusing stage. The interferometer was illuminated with a stabilized helium neon laser (wavelength 632.8 nm) and the signals from the reference and measurement mirrors were recorded continuously using an NPL developed field programmable gate array (FPGA) based fringe counting system. The double pass configuration meant that one fringe corresponded to a quarter of the wavelength (158 nm) rather than the usual half wavelength associated with a Michelson interferometer. To determine the stage displacement the phase change of the signals from the reference and measurement mirrors was measured and a Heydemann correction was applied to remove interferometer non-linearity [17,18]. Real-time values for the relative displacement were then calculated as the focusing stage executed a preprogramed 'staircase' displacement sequence, pausing after each micrometre of travel for one second to collect data at each position. The total calibrated travel range of ~150 µm was chosen to encompass the axial range used to image the calibration pattern.

The transverse magnification of the microscope was calibrated by imaging in reflection a silicon calibration standard used in atomic force microscopy. The standard comprised of a regular array of square pillars with a 3 µm pitch, which was calibrated using optical diffractometry. Any scaling differences between images of the pitch standard at the illumination wavelength and the emission wavelengths of the fluorescent sample are assumed to be negligible.

When measuring the axial separation of the features, additional depth corrections will be required to account for differences in objective lens and immersion media between fabrication and imaging setups. However, as long as the fabrication parameters, refractive index and dispersion properties of the polymer are known, corrections to the measured separation can be made for any choice of imaging objective and immersion media. For the fluorescent calibration patterns, the effects of aberration arising from refractive index mismatch between sample (n=1.58) and immersion medium (n=1.51) are minimised by having a small (<20 µm) separation between the fabricated pattern and the top surface of the polymer layer.

*2.4 Determining lateral resolution from fluorescent features*

There are three methods by which it is possible to determine lateral resolution from images of the fluorescent feature. The first makes use of the thickness of the fluorescent shell and is based on the assumption that the fluorescent shell thickness is small compared to the PSF width. A fluorescence image of the feature is then the optical section of the fluorescent shell convolved with the microscope PSF. A more accurate estimate of the PSF width can then be obtained from the fluorescent shell image by applying the known thickness and diameter of the fluorescent shell to a standard deconvolution algorithm.

A second method is to take advantage of the controllable shell diameter to fabricate rows of features whereby the shell diameter decreases with row number. A quick determination of the image resolution limit is to identify the row for which there is less than a 26% dip in intensity at the centre of the image of the feature (Rayleigh criterion). The shell diameter at this point provides an estimate of the imaging resolution. This will yield the same estimate as that obtained from the shell width alone. Again, as the shell has a finite width, without deconvolution this method will provide the same overestimate of PSF width as the first method. A third and final method is to simply make the shell diameter as small as possible and treat the entire feature as a sub-diffraction bead.

Without deconvolution, the resolution estimate provided by these methods would be no worse than imaging a bead with a diameter equal to the shell width. As the volume of the fluorescent shell becomes much larger than the void volume below 200 nm, laser written fluorescence offers a significant advantage in signal-to-noise over beads when determining PSF width in super-resolution microscopes. Another added benefit is the ability to locate the fluorescent features on a regular grid across the field of view to systematically map out field dependent changes in PSF width. A further advantage of this approach is that the circular symmetry of each shell avoids any rotational dependence of the reported resolution, as seen for some 1D gratings in laser scanning microscopes [19].

*2.5 Distortion characterisation*

Distortion is a spatially dependent magnification change and is typically manifested as either a pincushion or barrel distortion of the microscope image. Correction of distortion is achieved by treating the fluorescent grid pattern (Fig. 1(b)) as the product of two orthogonal 1-dimensional gratings (X-grating and Y-grating). Distortion is interpreted as a local change in the phase of the 1D grating. The phase can be extracted after spatial filtering of the first (+1) diffraction order associated with the grating. The spatially filtered diffraction order is inverse transformed and the phase in this plane is calculated for each point in the field. As the distortion function is assumed to be slowly varying (low spatial frequency), unwrapping the phase is relatively straightforward. The phase ramp associated with the grating periodicity is subtracted from the unwrapped phase to obtain two phase shift maps (X and Y). This phase shift map is converted into a pixel shift map by dividing by the spatial frequency of the grating in radians per pixel. Finally, the X and Y pixel shift maps are used together with the *imwarp* function (MATLAB Image Processing Toolbox) to calculate the distortion-corrected image.

## 3. Results

### 3.1 Spatial accuracy and precision

The array of 8 × 8 × 3 features was imaged on a laser scanning confocal microscope build around an Olympus IX81. A 60X 1.35 NA objective was used to focus 488 nm laser excitation onto the sample with fluorescent emission collected between 500 nm and 600 nm. During data acquisition, the difference between the axial displacement indicated by the stage encoders and that measured by the interferometer was between -0.6 μm and +0.6 μm. A 6th order polynomial was fit (root mean square of residuals 92 nm) to these discrete measurements to obtain a continuous stage error curve which was used to correct the nominal axial separation of image planes in subsequent focal stacks of the fluorescent feature array.

Lateral and axial (XZ) sections through the acquired 3D data set is shown in Fig. 2. The coordinates (X Y Z) of each fabrication feature in the array was calculated using the centre of mass (centroid) of the fluorescence intensity data associated with each feature. A 20 slice Z-stack, containing one of the 8 × 8 arrays, was first selected. Within this truncated Z-stack, a square tile (side length 1.7 μm) was used to segment the region around each of the 64 features. The total fluorescence within the tile area was summed for each image in the stack and plotted as a function of depth. The centroid of this intensity profile was calculated, taking into account the Z value corrections provided by the interferometer. The lateral (XY) centroid was then calculated for the tile at a depth corresponding to the Z centroid. This process was repeated for each of the three 8 × 8 arrays.

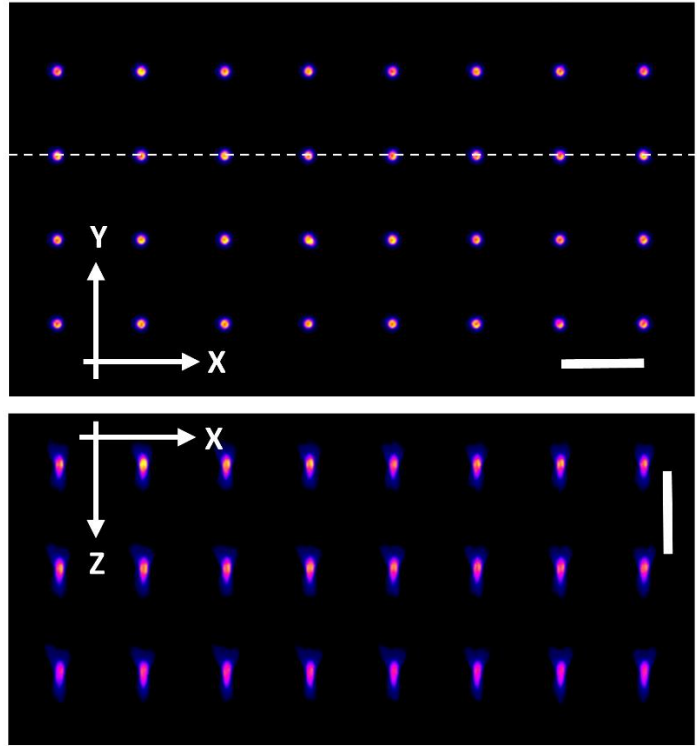

Fig. 2. XY section through the confocal data stack of the 8 × 8 × 3 array. The dashed line indicates the location of the XZ section shown below. Z increases with distance into the sample. Scale bars: 10 μm.

With XYZ coordinates for the centroid of each fluorescent feature, the axial and lateral separations could be accurately determined. The results of the mean lateral (ΔX, ΔY) and axial (ΔZ) separations between neighbouring features is shown together with standard deviations (σ) in Table 1. For each sample there were 7 × 7 × 3 = 147 measures of lateral separation (ΔX, ΔY) and 8 × 8 × 2 = 128 measures of the axial separation (ΔZ). All values are in μm.

Table 1. Mean and standard deviation values for feature separation (in μm)

| Sample No. | <ΔX>  | <ΔY>   | <ΔZ>   | σ(ΔX) | σ(ΔY) | σ(ΔZ) |
|---|---|---|---|---|---|---|
| 1 | 9.971 | 10.012 | 9.923  | 0.037 | 0.029 | 0.110 |
| 2 | 9.971 | 10.014 | 10.144 | 0.024 | 0.026 | 0.087 |
| 3 | 9.980 | 10.017 | 10.012 | 0.027 | 0.021 | 0.100 |

*3.2 SIM lateral resolution and shell thickness*

To demonstrate that the shell thickness is independent of the feature size, features were created using a range of laser powers (Fig. 3). The features were then imaged using a custom-built structured illumination microscope, developed at the National Physical Laboratory [20,21]. Transects taken through the images of each feature size are shown in Fig. 5. Each profile shown is an average of three line profiles taken from features fabricated with the same pulse energy and pulse number. The FWHM averaged across all transects was 208 ± 15 nm (N=6). The variation in shell thickness and shell diameter with fabrication parameters is shown in Fig. 4.

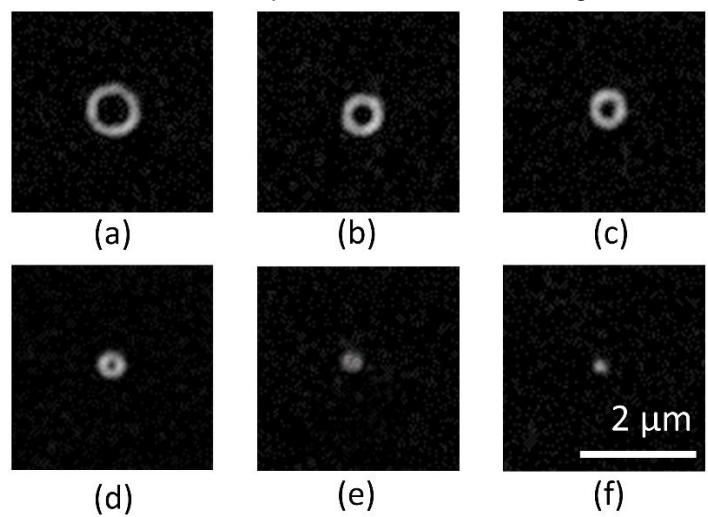

Fig. 3. SIM images of individual features fabricated using a range of pulse energies and repetitions. (a)-(c): five pulses, with energies of: 8 nJ (a), 4.7 nJ (b) and 3.8 nJ (c). (d)-(f) single pulse, with energies of: 5.7 nJ (d), 4.7 nJ (e), 3.8 nJ (f). Image brightness has been adjusted for clarity

Estimates of the fluorescent shell thickness can be obtained using images of beads made on the same system. The FWHM of an image of a 100 nm fluorescent bead (YG Fluospheres, ThermoFisher) was measured to be 170 ± 13 nm. The bead image was used as a model for the system PSF to deconvolve SIM images of an 8 × 8 array of features. By taking line profiles through the features in the deconvolved image, the average value measured for the shell thickness was 128 ± 19 nm (N=10).

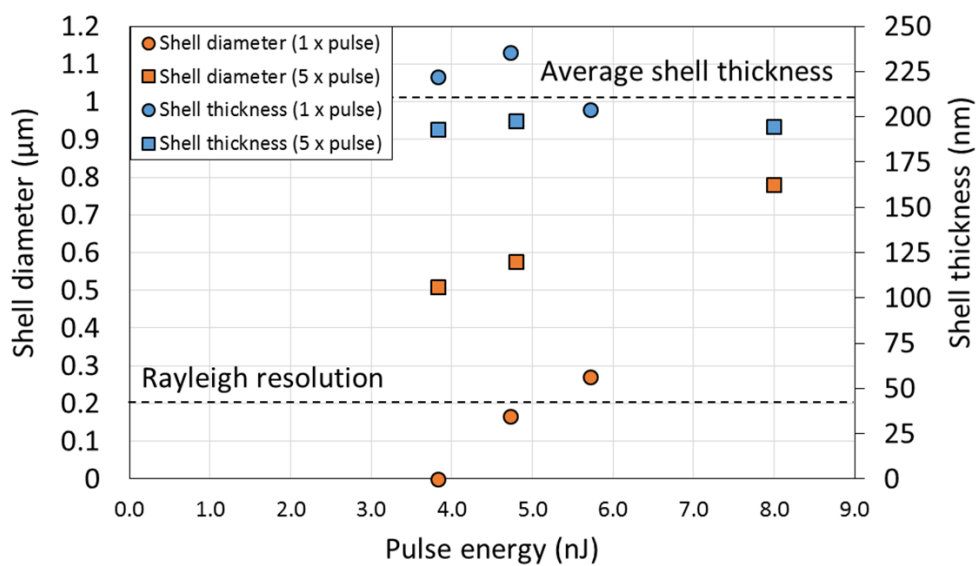

Fig. 4. Relationship between shell diameter, shell thickness and fabrication parameters for the SIM images shown in Fig. 3. The plots indicate a linear relationship between shell diameter and pulse energy, whilst the apparent shell thickness remains independent of pulse energy.

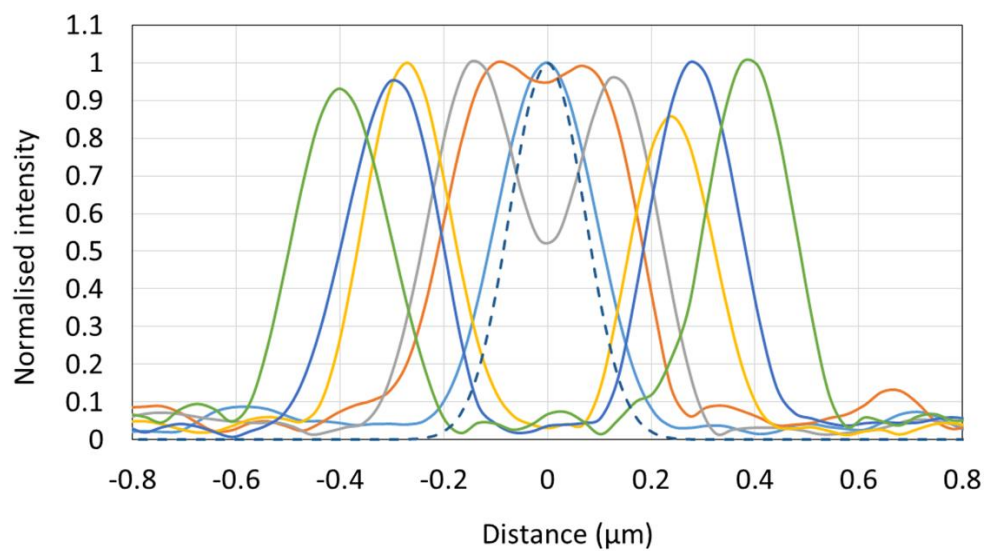

Fig. 5. Line transects across SIM images of the fluorescent shells shown in Fig. 3. The line profiles correspond to features fabricated with the following parameters: 5 × 8 nJ (green) 5 × 4.7 nJ (blue) 5 × 3.8 nJ (yellow), 1 × 5.7 nJ (grey), 1 × 4.7 nJ (orange), 1 × 3.8 nJ (blue). The dashed blue line shown the profile of a 100 nm fluorescent bead captured on the same microscope.

### 3.3 Measuring confocal resolution

The ability of fluorescent features to determine microscope resolution was tested by imaging features of variable diameter on a Zeiss Airyscan confocal microscope. Images of the features are shown in Fig. 6. Transects taken through the images of each feature size are shown in Fig. 7, each profile representing the average of three line profiles taken from different features fabricated with the same pulse energy and pulse number. As discussed above, several methods were proposed for measuring the microscope resolution. The first method determined whether the size of the intensity dip at the centre of the feature was at least 26% of the peak signal value. The last shell diameter for which this condition holds then provides an upper limit for the resolution estimate, in this case, 450 nm. The first shell diameter which did not meet the Rayleigh criterion had a shell dimeter of 327 nm. The resolution is then determined to be between these two limits.

The second method for determining lateral resolution involved measuring the FWHM of the shell thickness. For cases where the shell diameter was too small to measure the FWHM, a value equal to twice the half-width at half maximum was used. Using this approach, the average FWHM across all transects was 338 ± 23 nm (N=6), which is consistent with the range predicted by the first method. These estimates can be compared to the width of 303 ± 10 nm (N=5) measured on the same system using 100 nm beads (YG Fluospheres, ThermoFisher).

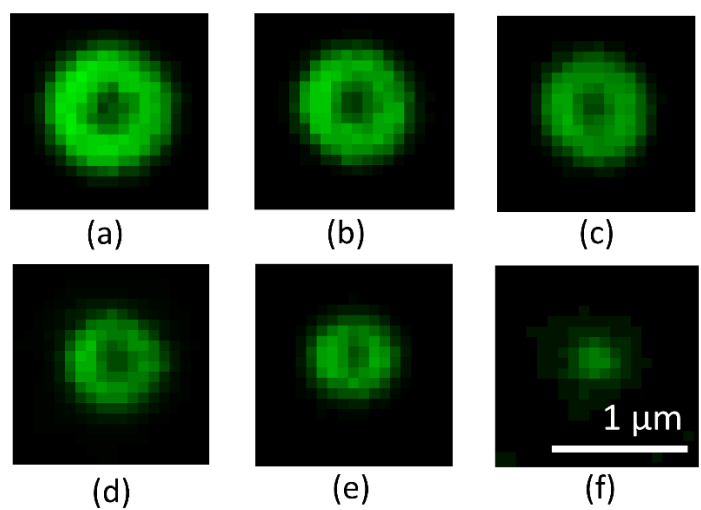

Fig. 6. Images of fluorescent features taken on a Zeiss Airyscan confocal microscope operating in standard confocal mode. (a)-(d): five pulses, with energies of: 7.4 nJ (a), 6 nJ (b) 5.4 nJ (c) and 4.8 nJ (e)-(f) single pulse, with energies of: 7.4 nJ (e), 4.2 nJ (f). Image brightness has been adjusted for clarity.

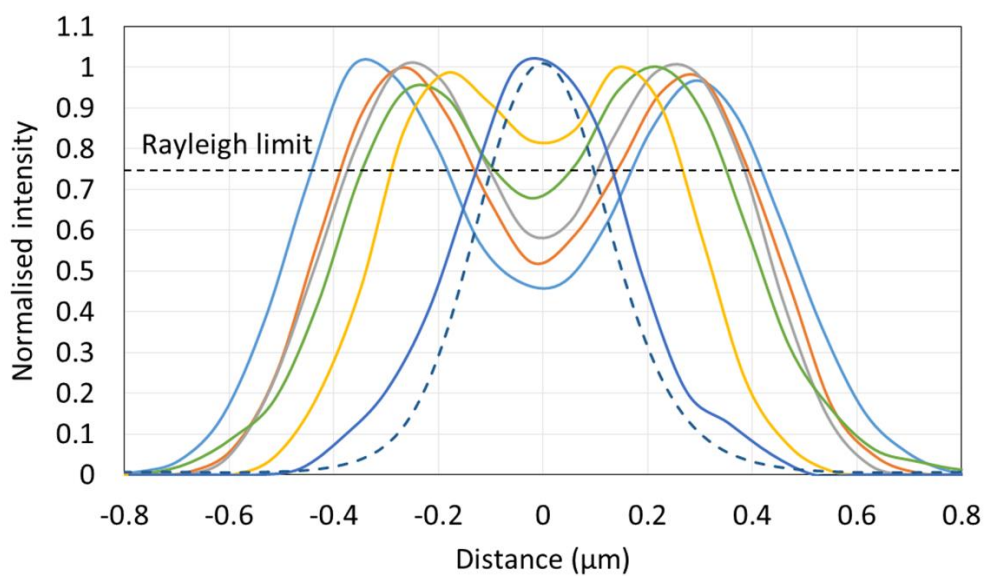

Fig. 7. Line transects through fluorescent shells shown in Fig. 6. The dashed blue line is an average line profile through images of a 100 nm bead taken on the same microscope. The black dashed line indicates the Rayleigh limit which determines the condition for the separation of the two sides of the fluorescent shell to be distinguishable.

*3.4 Distortion correction and illumination uniformity*

The application of the grid pattern to calibrate distortion and illumination uniformity is illustrated in Fig. 8. The grid image was captured on a widefield fluorescence microscope (Zeiss Axioplan 2) using a 40X 0.6NA Zeiss LD Plan-NEOFLUAR objective lens. The image in Fig. 8(a) clearly shows a bright hot spot in the centre caused by imperfect source alignment, as well as a significant pincushion distortion. The illumination profile is obtained from the grid image by taking the Fourier transform and applying a low pass spatial filter (red circle in Fig. 8(b)). The circular spatial filter has a radius equal to half of the grid spatial frequency. This choice of filter radius provides a balance between delivering a smoothly varying intensity map and sufficient spatial frequencies to clearly describe localised hot spots. The illumination non-uniformity recovered from the inverse Fourier transform is shown in Fig. 8(c).

Whilst the illumination uniformity could have been obtained from any uniform fluorescent sample, the choice of a fluorescent grid offers the advantage that the same image can also be used to measure image distortion. The first order in the 'X' direction is spatially filtered (green circle in Fig. 8(b)) and inverse transformed. The phase profile is then recovered from this complex image (Fig. 8(d)) and unwrapped. The difference between this phase profile and the ideal linear phase ramp is calculated to obtain the distortion. Phase distortion can be transformed into pixel shifts using the average grating period in pixels. The process is then applied to the 'Y' first diffraction order. The X and Y pixel shift values are used as input to the MATLAB *imwarp* function to perform the inverse distortion.

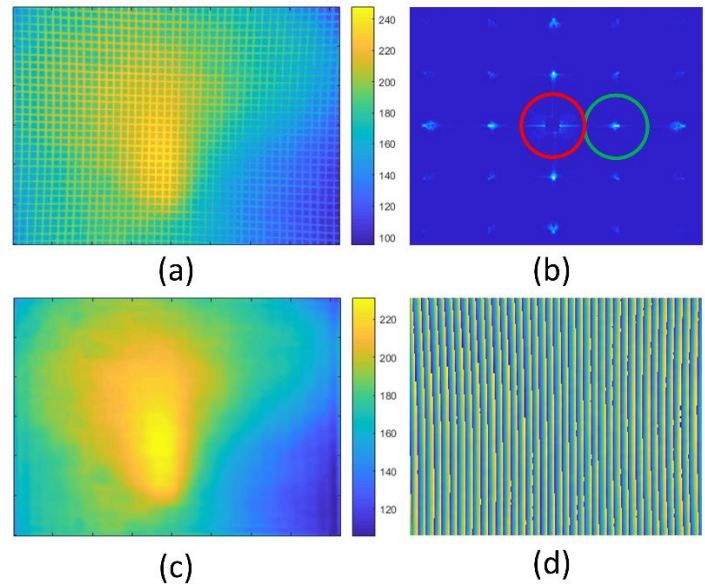

Fig. 8. (a) Raw image of the fluorescent grid using a 40X objective lens taken on a Zeiss Axioplan 2. (b) The Fourier transform of (a) with the zero order blocked to better visualise location of higher orders. (c) Inverse Fourier transform of region passing through red spatial filter in (b). (d) The wrapped phase angle of the signal passing through the green spatial filer in (b). Scale bar: 100 μm.

To demonstrate the accuracy achievable in the distortion correction algorithm, it was first tested on synthetic data. An ideal grid (Fig. 9A) with a 10-pixel grid spacing and 2 pixel line width was warped using the same distortion pattern as measured for the Zeiss 40X 0.6NA lens (Fig. 8). This warped image was then used as the input image for the distortion correction algorithm.

The output image from the distortion correction algorithm (shown in red in Fig. 9(b)) was scaled and overlaid with the original ideal image (shown in white). It can be seen that the only evidence of the red un-warped grid underneath that of the white 'perfect' grid is a few resampling artefacts, indicating sub-pixel precision (i.e. less than 1 μm). An image showing the size and direction of the pixel shifts required to correct the distortion is shown in Fig. 9(c), with numbered regions shown magnified in Fig. 9(d).

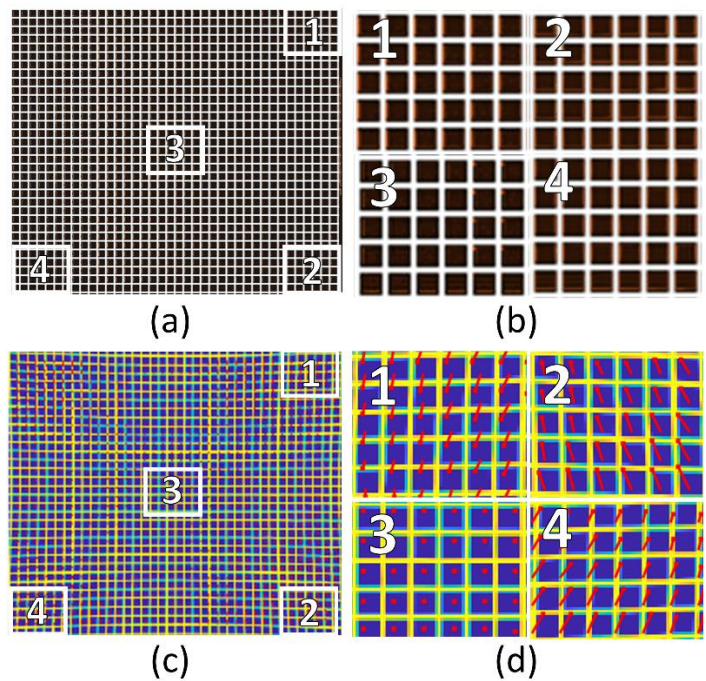

Fig. 9. Demonstrating the accuracy of the distortion correction algorithm. (a): An ideal grid was warped using the MATLAB imwarp function before being used as input into the distortion correction algorithm. (b) Details from the overlap of the ideal grid (white) on the distortion corrected grid (red). Each tile corresponds to the numbered regions highlighted in (a). (c) Pixel shift map showing the size and direction of the unwarping within each region to recover the original grid pattern. (d) Details of the pixel shift map, with tiles corresponding to the numbered regions in (c).

*3.5 Colour plane alignment*

A final example of the utility of the fluorescent calibration patterns is in colour plane alignment. Non-common path errors can introduce displacements between colour planes which is critical to multi-colour applications e.g. in single molecule localisation [22]. Images of a single layer of the 8 × 8 × 3 array were taken on an Olympus FV3000 (Micron Advanced Bioimaging Unit in the Department of Biosciences, University of Oxford) (Fig. 10). Three excitation and detection bands were used to demonstrate the broadband response. The excitation wavelengths were 405 nm, 488 nm and 561 nm (Fig. 10(a)-(c)). The detection bands were 430-470 nm, 500-540 nm and 570-620 nm respectively. An image obtained by fusing each of the colour bands together into a single image is shown in Fig. 10(d). In this instance the colour plane misalignment was determined to be less than one pixel.

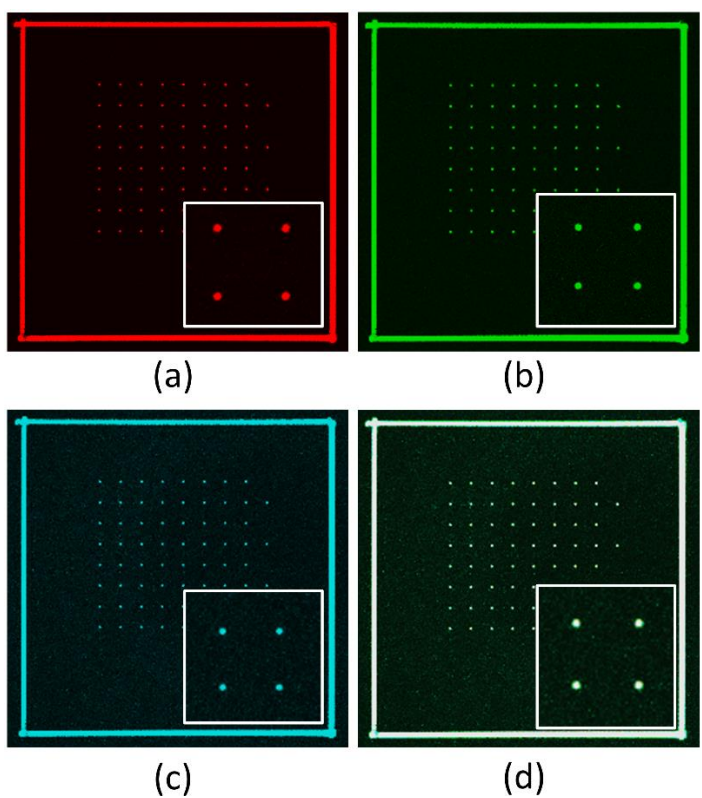

Fig. 10. Multi-channel images of a single layer of the 8 × 8 × 3 array, with a 2 × 2 detail shown inset. Excitation wavelengths (a)-(c) are: 405 nm, 488 nm and 561 nm, with detection bandwidths 430-470 nm, 500-540 nm and 570-620 nm respectively. (d) An image fusing the different colour planes into a single image showing the sub-pixel colour alignment. Images were acquired on an Olympus FV3000.

## 4. Conclusion

We have demonstrated a quick and easy to implement technique for the accurate determination of lateral resolution and image distortion using patterns generated by laser written fluorescence. To validate the spatial accuracy and precision with which it was possible to generate individual features, we developed a protocol that permitted the location of each feature to be determined with an accuracy of less than 30 nm in XY and less than 150 nm in Z. Applying this protocol we have measured the average spread (1σ values) in the feature location to be 29 nm in X, 26 nm in Y and 99 nm in Z.

Moreover, we have exploited the fluorescent shell structure created around the voids to estimate the lateral spatial resolution for structured illumination and confocal microscope images. The bright fluorescent shell around a dark void resembles the inverse of the traditional fluorescent bead used in microscope calibration. This presents a distinct advantage of retaining a high signal-to-noise ratio as the shell diameter approaches zero, providing a reliable means of measuring the resolving power of super-resolution microscopes. Using measurements of 100 nm fluorescent beads to estimate the SIM PSF, the deconvolved thickness of the fluorescent shells was measured to be 128 ± 19 nm. With adjustments for immersion media and objective type, any microscope can use the regularly spaced fluorescent features to calibrate the axial as well as lateral uniformity of the imaging volume for custom built and commercial microscopes.

A single image of a two-dimensional fluorescent grid pattern was used to quantify and correct both illumination non-uniformity and image distortion. The image distortion was corrected to better than 1 µm over a 400 µm field. If this information is captured immediately before or after imaging a sample, the calibration data can be used to retrospectively correct image data and thereby ensure well-defined spatial resolution, spatial uniformity and intensity uniformity.

Using laser written fluorescence, it is anticipated that further fluorescent patterns will be developed to provide a wider range of imaging performance parameters, such as a single shot measure of sectioning thickness or to enable aberration correction across three-dimensional volumes.

**Funding**

Open Innovation Proof of Concept Award (University of Exeter)

University Challenge Seed Fund (University of Oxford) (UCSF 427)

Wellcome Trust Strategic Award 107457 (University of Oxford)